\documentclass[fleqn,usenatbib]{mnras}

\usepackage{newtxtext,newtxmath}

\usepackage[T1]{fontenc}
\usepackage{ae,aecompl}
\pdfoutput=1

\usepackage{graphicx}	
\usepackage{amsmath}	
\usepackage{amssymb}	
\usepackage[space]{grffile}
\usepackage{latexsym}
\usepackage{textcomp}
\usepackage{longtable}
\usepackage{multirow,booktabs}
\usepackage{amsfonts,amsmath,amssymb}
\usepackage{url}
\usepackage{hyperref}
\hypersetup{colorlinks=false,pdfborder={0 0 0}}
\usepackage[utf8]{inputenc}
\usepackage[english]{babel}
\usepackage[]{graphicx}
\usepackage{pdflscape}
\usepackage[]{subfigure}
\usepackage[]{color}
\usepackage[]{bm}
\usepackage[]{natbib}

\newcommand{\simgt}%
        {\,\hbox{\lower0.6ex\hbox{$\sim$}\llap{\raise0.6ex\hbox{$>$}}}\,}
\def\simlt{\lower.5ex\hbox{\ltsima}}
\def\ltsima{$\; \buildrel < \over \sim \;$}
\def\simlt{\lower.5ex\hbox{\ltsima}}
\def\gtsima{$\; \buildrel > \over \sim \;$}
\def\simgt{\lower.5ex\hbox{\gtsima}}

\def\arcsec{\hbox{$^{\prime\prime}$}}

\newcommand{\msun}{\mbox{M$_\odot$}}

\title[Magnetically regulated collapse in B335]{Magnetically regulated collapse in the B335 protostar ?
\\ I. ALMA observations of the polarized dust emission}

\author[A. J. Maury et al.]{
A. J. Maury$^{1,2}$\thanks{E-mail: anaelle.maury@cea.fr},
J. M. Girart$^{3,4}$,
Q. Zhang$^{2}$, 
P. Hennebelle$^{1}$,
E. Keto$^{2}$,
R. Rao$^{5}$,
\newauthor
S.-P. Lai$^{6}$,
N. Ohashi$^{5,7}$,
and M. Galametz$^{1}$
\\
$^{1}$Laboratoire AIM, CEA/DSM-CNRS-Universit\'e Paris Diderot, IRFU, Astrophysics department, 91191 Gif-sur-Yvette, France\\
$^{2}$Harvard-Smithsonian Center for Astrophysics, Cambridge, MA 02138, USA\\
$^{3}$Institut de Ci\`encies de l’Espai (ICE, CSIC), Campus UAB, E-08193 Cerdanyola del Vall\`es, Catalonia, Spain\\
$^{4}$Institut d'Estudis Espacials de Catalunya (IEEC), Catalonia, Spain\\
$^{5}$Institute of Astronomy and Astrophysics, Academia Sinica, 645 N. Aohoku Pl., Hilo, HI 96720, USA\\
$^{6}$National Tsing Hua University, Taiwan \\
$^{7}$Subaru Telescope, National Astronomical Observatory of Japan, 650 N Aohoku Pl., Hilo HI 96720, USA
}

\date{Accepted 2018. Received 2018; in original form 2018}

\pubyear{2018}

\begin{document}
\label{firstpage}
\pagerange{\pageref{firstpage}--\pageref{lastpage}}
\maketitle

\begin{abstract}
The role of the magnetic field during protostellar collapse is poorly constrained from an observational point of view, although it could be significant if we believe state-of-the-art models of protostellar formation. We present polarimetric observations of the 233\,GHz thermal dust continuum emission obtained with ALMA in the B335 Class~0 protostar.  Linearly polarized dust emission arising from the circumstellar material in the envelope of B335 is detected at all scales probed by our observations, from radii of 50 to 1000 au.
The magnetic field structure producing the dust polarization has a very ordered topology in the inner envelope, with a transition from a large-scale poloidal magnetic field, in the outflow direction, to strongly pinched in the equatorial direction. This is probably due to magnetic field lines being dragged along the dominating infall direction since B335 does not exhibit prominent rotation.
Our data and their qualitative comparison to a family of magnetized protostellar collapse models show that, during the magnetized collapse in B335, the magnetic field is maintaining a high level of organization from scales 1000 au to 50 au: this suggests the field is dynamically relevant and capable of influencing the typical outcome of protostellar collapse, such as regulating the disk size in B335.
\end{abstract}



%

\section{Introduction}

The recent development of numerical magneto-hydrodynamical models describing the collapse of protostellar cores and the formation of low-mass stars, has opened new ways to explore in more details the physical processes responsible for the characteristic outcomes from the formation of solar-type stars. Unfortunately, a major ingredient in these models, the magnetic field itself, still remains very poorly constrained from an observational point of view, while from the theoretical point of view, its role has been largely described \citep[e.g.][for a review]{LiPPVI}. 
For example the role of magnetic braking, if the field is strong enough and well coupled to the envelope material, in regulating disk formation during the Class~0 phase, has been the focus of recent studies \citep[][]{Hennebelle16,Krasnopolsky11,Machida11} because it could potentially explain the low-end of the size distribution of protostellar disks \citep[e.g.][]{Maury10, Maury18}.
Beyond understanding disk formation, characterizing the roles of the magnetic field during the main accretion phase is a long standing issue, possibly tied to several important open questions in the field of star formation, such as the launching of jets and outflows or even the magnetic flux problem for main-sequence stars.

When the dust grains are in an environment permeated by a magnetic field, their long axes partially align perpendicularly to the magnetic field lines \citep[][]{Lazarian07}. The thermal emission of the dust grains thus becomes polarized, with a polarization direction perpendicular to the magnetic field component projected onto the plane of the sky. Dust polarized emission is therefore the most widely used observational constraint used to trace the magnetic field in star-forming cores: while it was mostly limited to massive objects for sensitivity reasons before the ALMA era, the core-scale magnetic fields have been mapped in an handful of low-mass protostars \citep[e.g.][]{Girart06,Rao09,Hull14,Stephens13}, and several hourglass shapes (see section 4.4) were observed at core scales suggesting the field is well coupled to the infalling material at those scales.

B335 is an isolated Class 0 protostar with a bolometric luminosity of $\sim 1 L_{\sun}$, at a distance of 100 pc \citep{Olofsson09}, with an east-west elongated molecular outflow, nearly in the plane of the sky \citep[][]{Hirano88}, which dynamical timescale $3\times 10^{4}$ yrs \citep[][]{Stutz08} strengthens the scenario of a young central protostellar embryo. 
Infall motions have been detected in the B335 envelope with both single-dish and interferometric observations of
molecular emission lines, and models to derive mass infall rate have been attempted 
\citep[][]{Zhou93,Saito99,Evans05,Evans15,Yen10, Yen15b}. 
The modeled values suffer from large uncertainties, with mass infall rates ranging from  $10^{-7}$ \msun yr$^{-1}$  to $\sim 3 \times 10^{-6}$ \msun yr$^{-1}$ at radii of 100--2000 au, and infall velocities from 1.5 km s$^{-1}$ to $0.8\pm0.2$ km s$^{-1}$ at $\sim 100$ au. Hence the central protostellar mass is subject to high uncertainties, in the range $\sim$ 0.04-0.15 \msun. 
ALMA sub-millimeter dust emission shows that B335 remains single on scales $\sim 100$ au \citep[][]{Evans15, Imai16}. The mass derived from dust emission using the optically thin assumption inside a radius 25 au is $<10^{-3}$\msun, while the mass of accreted material (mass infall rate integrated over the outflow timescale) is $\sim0.15$\msun. Hence, most of the material is either in the envelope or has been accreted onto the central protostar, and not much mass could be residing in an optically-thin disk.
\\Kinematics of the B335 core have been studied from $\sim20,000$ to $\sim10$ au, showing a slowly rotating outer envelope \citep[][]{Saito99,Yen10} and suggesting a braking of the rotation in the inner core because the radial profile of the specific angular momentum flattens at $\sim1000$ au \citep[][]{Kurono13} and no clear rotational motion is detected at tens of au scales \citep[][]{Yen15b}.
All these observations suggest that any protostellar disk in the B335 protostar is smaller than what would be expected from simple conservation of the angular momentum, 
and most likely only present at scales $<<50$ au.
\\Because of its youth, strong accretion and absence of identified rotational motions in the inner envelope, B335 is an excellent candidate for the magnetically-regulated protostellar collapse scenario. The first step to test whether the magnetic field can be responsible for the braking of the envelope is to characterize the magnetic field in the protostellar environment.
After detecting dust polarization in B335 with the SMA in the framework of our polarization survey of Class 0 protostars (Galametz et al. in prep), we carried out ALMA observations to obtain a high dynamic range characterization of the magnetic field topology in this young Class 0 protostar.

\begin{figure*}
   \centering
   \includegraphics[trim={0 0 0 0},clip,width=0.98\linewidth]{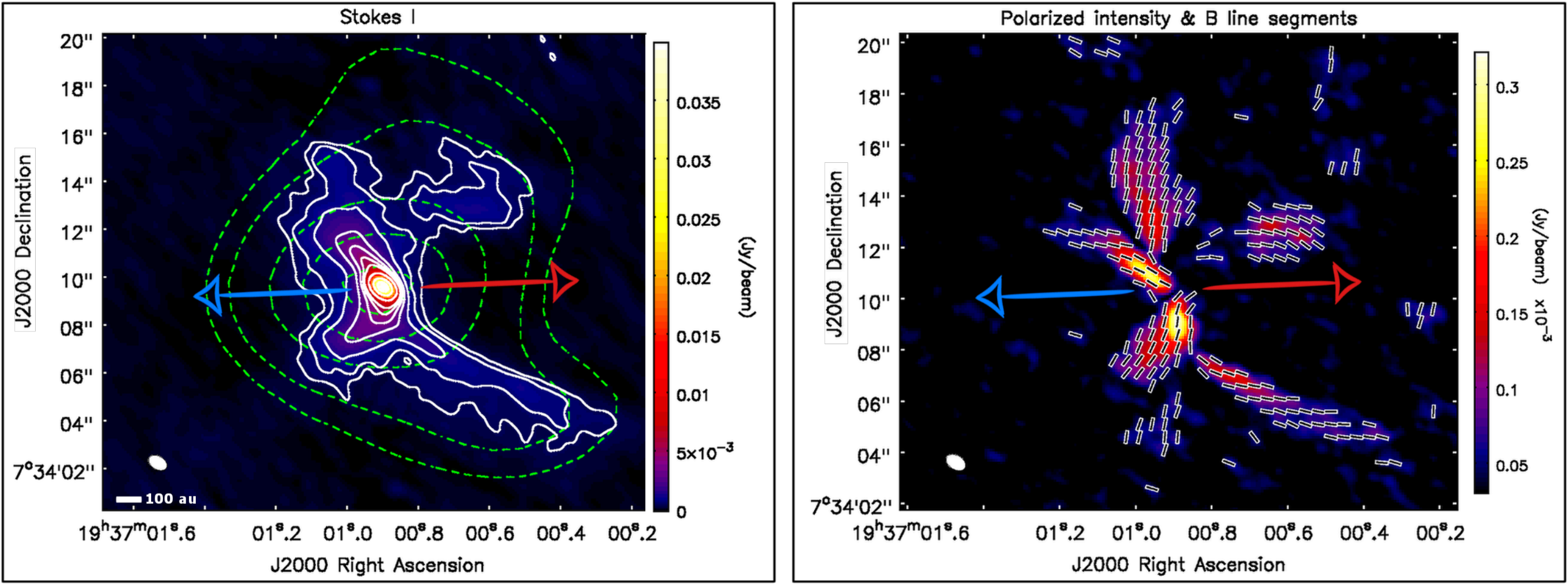}
      \caption{Left: The background map and white contours show levels of the 1.29\,mm
      Stokes I emission (rms 120 $\mu$Jy/beam, levels of 3, 5, 10, 20, 30, 40, 50, 100, 200, 300 $\sigma$). The green dashed contours show levels of dust continuum emission at the core scale, as traced by the ACA dataset from the ALMA archive (project 2013.1.00211.S, PI D. Mardones). Right: The background image shows the polarized dust continuum emission. 
      The superimposed line segments show the B-field (polarization angle rotated by $90^{\circ}$) where the polarized dust continuum emission is detected at $>3\sigma$.
              }
         \label{Fig-StokesI-polar}
\end{figure*}

\section{Observations and data reduction}

B335 was observed with the ALMA interferometer on 2015-05-30, with 34 antennas  sampling baselines from 21m to 639m (peak of baseline density around 200m), under good atmospheric conditions (PWV$\sim$ 0.8 mm).
The receiver was tuned to cover the 223-227 GHz and 239-243 GHz windows, using an aggregated bandwidth 7.5 GHz for the continuum emission (4 basebands of 1.87 GHz bandwidth each), in full polarization mode. The average system temperatures were $\sim$65-70 K. The phase center was $\alpha$(J2000) = $19^{\rmn{h}} 37^{\rmn{m}} 00\fs91$, $\delta$(J2000) = $+07\degr 34\arcmin 09\farcs 6$.
The gain calibrators used were the QSO J1955+1358 and J1751+0939. 
J1751+0939 was used for the polarization calibration: the parallactic angle coverage of this QSO span over $100^{\circ}$.
The absolute flux scale was determined from observations of Titan, with an absolute flux uncertainty of $\simlt$ 10\%. 
The total time spent on the source was 74 minutes.
The data were calibrated by the NA Arc Node, and released to us on November 22, 2016.
Self-calibration of the phases was performed on the Stokes I dataset.
The maps have a synthesized beam $0.8\arcsec \times 0.54\arcsec$ ($\sim 80\times54$au), the rms noise in the Stokes I map is 0.12~mJy/beam while the Stokes Q and Stokes U map have rms noises $\sim$ 0.02 mJy/beam. The higher rms noise in the Stokes I map is due to limited dynamic range.
Analysis of the scatter in the parallel hands gain after polarization calibration of the calibrators indicates an instrumental polarization < 0.5\%.

\section{ALMA polarized dust emission map in B335}

Our ALMA Stokes I map of the 1.29\,mm dust continuum emission in B335 has a spatial resolution $\sim 60$\,au: the maximum emission is found at (19:37:00.897;07:34:09.592) with a peak flux density 44 mJy/beam. The map, sensitive to the dust continuum structures emitting at scales from $\sim 60$\,au to $\sim 1000$\,au, is shown in the left panel of Figure \ref{Fig-StokesI-polar}. It strikingly traces mostly two structures: (i) the north-south high density inner envelope around the central protostar, and (2) the east-west outflow cavity walls.
Polarized intensity and polarization vectors maps are obtained from the combination of the Stokes Q and U maps, using only pixels where the Stokes Q and U fluxes exceeds 3 times the standard deviation.
The fractional polarization is found to be $\sim$4-11\% on average in the map, with a maximum of the polarized intensity 320 $\mu$Jy/beam at (19:37:00.892; 07:34:09.098), closely associated to the peak of the Stokes I dust continuum emission. Overall, the polarized millimeter continuum emission (background colorscale in the right panel of Fig. \ref{Fig-StokesI-polar} and contours in the right panel of Fig. \ref{Fig-Bangles}) follows closely the dust continuum emission. 
The magnetic field direction\footnote{Note that it is unlikely the polarization detected in B335 comes from self-scattering as was suggested by some ALMA observations of more evolved disks \citep{Kataoka15,Yang16} since densities to reach high scattering optical depth values that would produce high scattering polarization (especially the high polarization fraction values observed in our map since self-scattering is expected to yield $\sim 1\%$ of polarization fraction) are much higher than the gas densities probed at these scales in B335.} in the plane of the sky (polarization angle rotated by $90^{\circ}$ where the polarized intensity is detected at $>3\sigma$), is overlaid as white/black line segments in the central panel. 

\section{Discussion}

\subsection{Magnetic field topology in B335}

Interferometric filtering picks up the most compact and strongly polarized structures, hence the high density equatorial plane is preferentially traced by both the Stokes I and the polarized dust continuum emission.
The detection of strongly polarized emission originating from the outflow cavity walls is more puzzling.
We propose that it can be explained by an efficient dust grain alignment, favored along the cavity walls because of stronger illumination from the central protostar photons propagating in the cavity (Radiative Torque Alignment, \citealt{Lazarian07,Hull16}). Strong irradiation of the B335 outflow cavity walls is also supported by strong near-infrared emission along them (see {\it{Spitzer}} map in Figure 9 of Stutz et al. 2008), and their enhanced chemistry (molecular lines observed in our Cycle 4 program, in prep.).
The detection of polarized emission implies a high degree of organization of the B field locally and along the line of sight: picking up only the polarized emission from the cavity wall surface and the equatorial plane also suggests a highly organized field in these two areas despite their different local properties (density, irradiation).

\begin{figure*}
    \centering
    \includegraphics[width=0.98\linewidth]{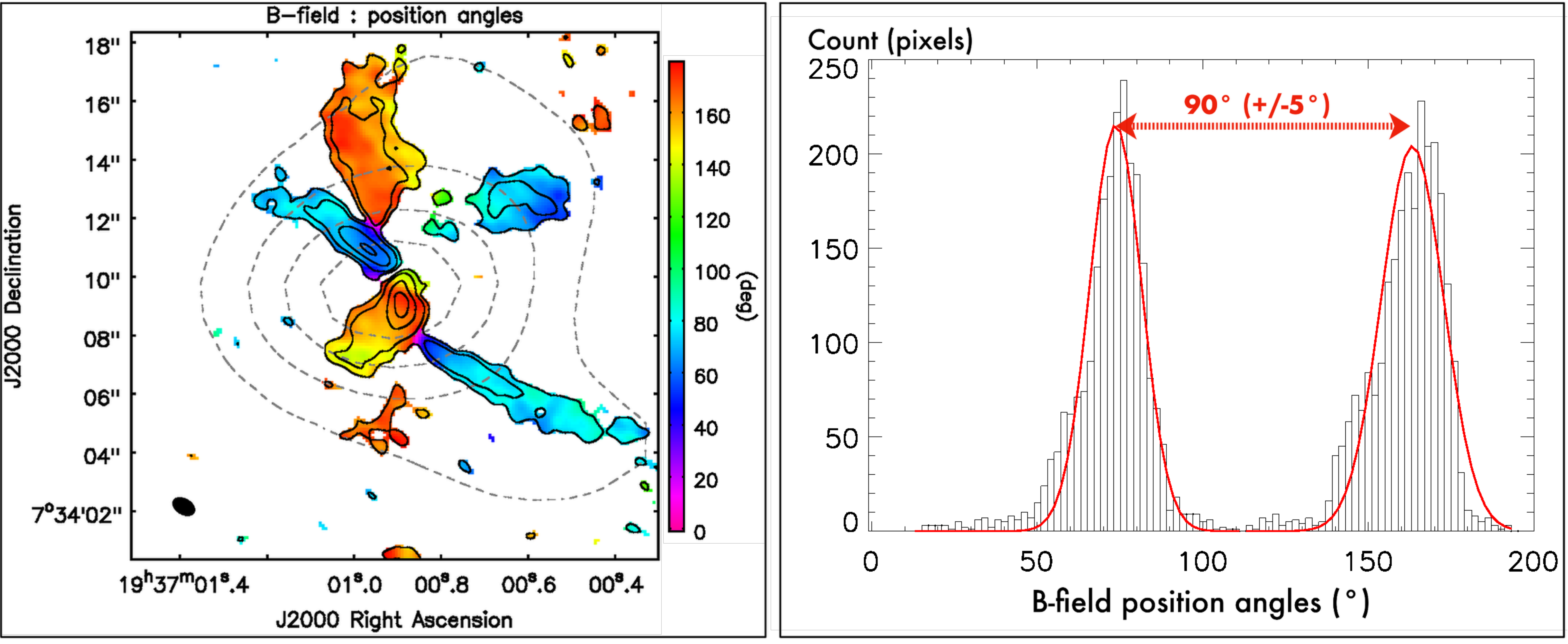}
       \caption{Left: The background image shows the map of position angles for the magnetic field line segments, computed only where polarized dust continuum emission is detected with $P_{\rm l}/\sigma(P_{\rm l}) > 3$. The black contours show polarized intensity levels (3,5,10,$15\sigma$) and the gray dashed contours show the dust continuum emission from the 7m-array data probing the core scales (as in Figure \ref{Fig-StokesI-polar}). Right: the histogram shows the distribution of the magnetic field position angles in the central $21\arcsec$ area of our ALMA map. The red lines show the Gaussian minimization to the histogram, used to compute the position and FWHM of the double-peaked distribution.
               }
          \label{Fig-Bangles}
\end{figure*}

The magnetic field topology traced in our ALMA map, shown in Figure \ref{Fig-StokesI-polar}, follows a very organized pattern, with the equatorial plane permeated by a field aligned north-south, while the outflow cavity walls trace a magnetic field mostly aligned along the outflow axis (note that there is also regions where both could be confused on the same line of sight because of the small viewing angle from a perfectly edge-on case for B335). 
In the right panel of Fig. \ref{Fig-Bangles} the histogram of the magnetic field position angles shows that the magnetic field in B335 has two directions largely prevailing at all scales 50-1000 au (two peaks at PA $74^{\circ}$ and $+164^{\circ}$).
Such a double-peaked distribution of B field line segments is consistent with a scenarii of an organized, strongly pinched magnetic field configuration \citep{Galli06} which large scales were filtered, as briefly discussed with the help of MHD models of protostellar collapse in section 4.3.
This scenario of strongly pinched magnetic field because of extreme stretching of the field lines along the inflow at the small scales probed here, deep in the potential well, is also supported by large scale observations of the magnetic field in B335: although they are affected by large uncertainties they suggest that the magnetic field orientation projected on the plane of the sky changes from mostly poloidal at 10,000 au (near-infrared polarization, \citealt{Bertrang14}) to mostly equatorial in the inner 1000 au (SCUBA, \citealt{Wolf03}).

\subsection{Magnetic field strength in B335}
Our ALMA data have enough statistics (>300 independent line segments) to attempt estimating the magnetic field strength from the polarization maps, using the statistical Davis-Chandrasekhar-Fermi method (\citealt{Davis51,Chandrasekhar53}, hereafter DCF method):
\begin{math}
|\vec{B}| = \sqrt{\frac{4\,\pi}{3} \, \rho_{\rm Gas}} \cdot \frac{\rm v_{\rm turb}}{\sigma_{\bar \theta}}\ 
\end{math}
, with $\rho_{\rm Gas}$ the gas density (g\,$\rm cm^{-3}$),
$\rm v_{\rm turb}$ the rms turbulence velocity (cm\,$\rm s^{-1}$), and
$\sigma_{\bar{\theta}}$ the standard deviation of the mean polarization angles (radians).
The standard deviation of the mean orientation angle $\sigma_{\bar{\theta}}$ should be estimated independently in each region of the hourglass to avoid contamination by the gravitational field (producing the two peaks separated by $90^{\circ}$), and compare only the turbulent field to the magnetic field, as intended in the DCF method. Here, we perform the DCF method using solely data in the equatorial region: we find $\sigma_{\bar{\theta}} \sim 0.38$ radians (note that a very similar value is found when using the cavity wall regions).
If we use $\rm{v}_{\rm turb}$ from 0.1 km$s^{-1}$ (similar to the thermal linewidth) to 1 km$s^{-1}$ (the typical broadening due to infall velocity measured at scales 100 au, \citealt{Evans15}), and a density profile in the dynamical free-fall region $\rho(r) = (c_{\rm s}^2/2\pi G)r^{-1.5}$ 
\citep{Kurono13}, we find a magnetic field strength 300-3000 $\mu$G, at radii 50-500 au.

Magnetic field estimates from the literature suggest the average field in the inner envelope is lower than the value we find: $130 \pm 40 \mu$G at 1000 au \citep{Wolf03}. Also, both our value and the inner envelope value of the magnetic strength are much larger than the field found at larger scales 12-40 $\mu$G \citep{Bertrang14}. 
If such a strengthening of the magnetic field, by almost two orders of magnitudes from scales 10,000 au to 100 au scales, is real (we stress that the validity of the DCF method in dense and dynamic environments remains to be validated), it could be explained by the strong dragging of magnetic field lines inward, as suggested from our ALMA B-field topology map at 50-500 au scales.

\subsection{Comparison to numerical models of magnetically-regulated collapse}

In order to test our hypothesis, we compared our ALMA magnetic line segments maps to non-ideal magneto-hydrodynamical numerical models of protostellar collapse \citep{Masson16} including ambipolar diffusion. 
Note that a quantitative comparison with the models is beyond the scope of this paper and will be presented in paper II (in prep.):
here we only aim at providing a first qualitative analysis from a simple morphological comparison to a family of models dedicated to reproduce the main observational feature of B335.  

Essentially we argue that the topology of magnetic field line segments found with ALMA is consistent with a 2 steps scenario. First a large scale magnetic field nearly parallel to the flow direction has been intensively pinched by the gravitational stress arising in the collapsing core to the point where the field lines run perpendicularly to the outflow in the equatorial plane. Second, the expansion of the magnetic cavities around the outflow is responsible for the field direction switching from being perpendicular to almost parallel to the outflow direction. Our models initial conditions feature a $2.5\msun$ core with $\beta=E_{rot}/E_{grav} \sim 10^{-2}-10^{-3}$, an initially uniform magnetic field of mass-to-flux ratio $\mu \sim 2-6$ and the angle between the rotation axis and magnetic field $\langle B,J \rangle = 10^{\circ}$.

\begin{figure}
   \centering
   \includegraphics[width=0.99\linewidth]{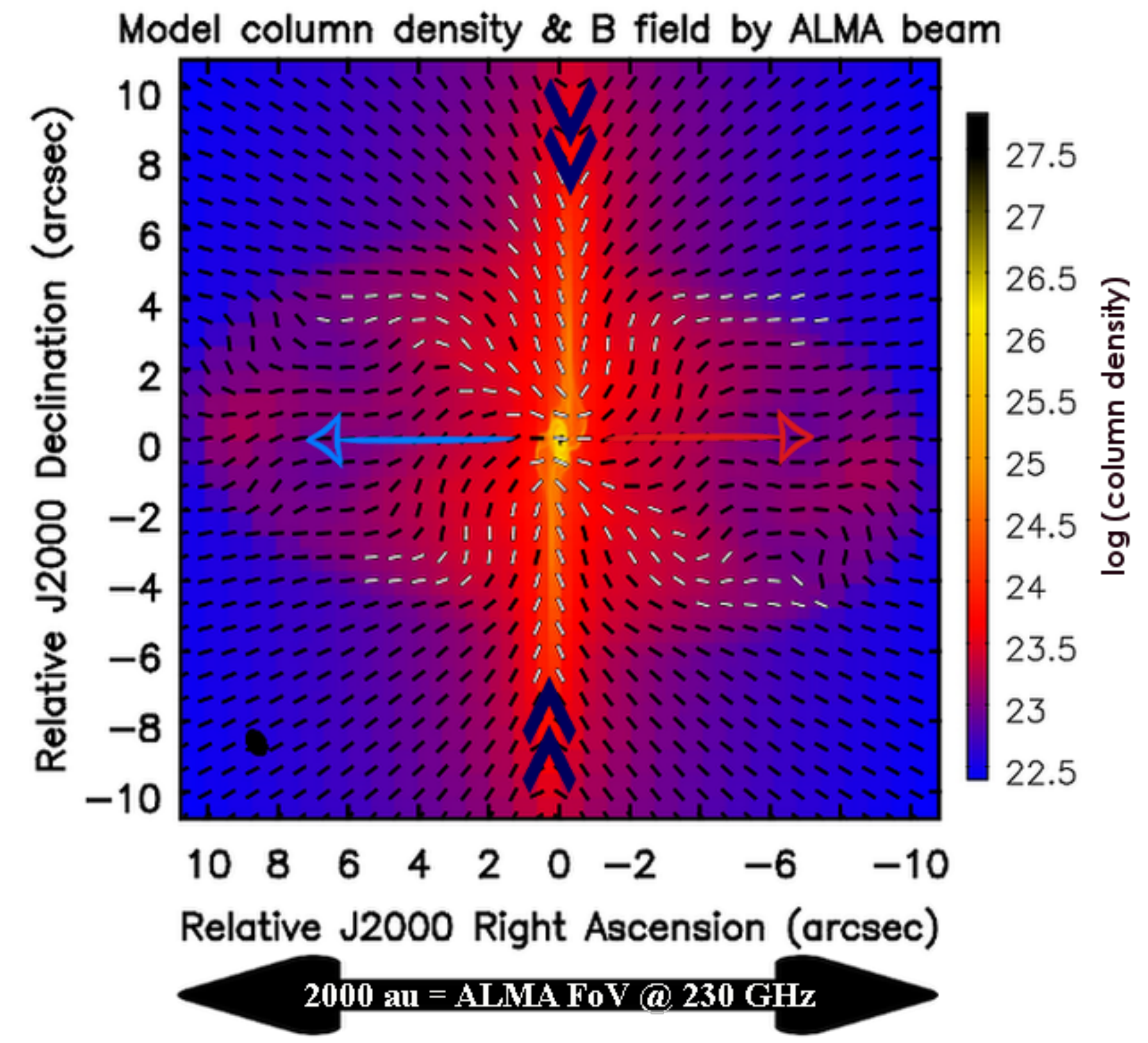}
      \caption{Non ideal MHD model of the protostellar collapse of a 2.5\msun ~core, showing the column density in the model, seen edge-on and projected at the B335 distance. The thick dark blue arrows indicate the main direction of the inflow, while the red/blue horizontal arrows indicate the direction of the outflowing gas. The magnetic field topology in the core at scales 2000 au, integrated along the line-of-sight and convolved by the ALMA synthesized beam, are shown as black line segments, while the white line segments highlight the general areas where they are detected in our ALMA map of B335.
              }
         \label{Fig-model-obs}
\end{figure}

Figure~\ref{Fig-model-obs} shows the plane-of-the-sky magnetic field line vector maps (integration along the line of sight weighted by the local density and convolved by a Gaussian pattern the size of the ALMA synthesized beam) from the best simple model found so far ($\mu=6$, $\beta=10^{-3}$), projected along an edge-on view (similar to the configuration for B335). The black line vectors highlight the B-field vectors where polarized dust emission is detected in our B335 map (either because of line-of-sight density effect or strong radiative alignment as shortly discussed in 4.1): it illustrates qualitatively that the magnetic topology obtained from the model is very similar to the observed one. We stress that a perfect agreement is not sought for here: for example the temperature effect is neglected when producing the model magnetic topology and hence the polarized contribution from the outflow cavities in the regions where they are on the line of sight is probably underestimated.

We emphasize that the pronounced pinching needed to explain the transition from an axial to an equatorial component as is observed in the B335 mid-plane is more easily obtained using MHD models with initial core mass-to-flux ratio $\mu > 5$. For $\mu<5$, the field is too strong and the bending of the field lines is not sufficient to reproduce the observed pattern.
Moreover, such an organized magnetic field topology is only obtained in models with a low level of initial turbulence, consistent with the small dispersion of the magnetic field position angles we observe in each region (histogram shown in Fig. \ref{Fig-Bangles}). Finally, not only this family of magnetically-regulated protostellar collapse models can reproduce the observed magnetic field pattern, but they also produce outflow cavities and small disks ($r<30$ au), consistent with the observed constraints on the centrifugal radius in B335 from molecular line observations. Although $\mu \geq 5$ is not a very strong field, it is still strong enough to be dynamically important, hence regulating the collapse such as affecting the disk formation, outflow launching and accretion rate, if compared to a hydrodynamical model with similar parameters.

\subsection{B335 in the context of other protostars}

The expected hourglass magnetic field morphology for the collapse of a slow or non-rotating sphere \citep[e.g.,][]{Galli93, Basu94} with an initially uniform field has been found in a handful of low and high mass dense cores \citep[e.g., IRAS 4A, G31.41, G240.31, NGC 6334, L1157:][]{Girart06, Girart09, Qiu14, Li-HB15, Stephens13}. Here, we discuss the two cases with a similar mass  and luminosities, IRAS 4A and L1157.  

For the case of IRAS 4A, the MHD models for a slightly supercritical ($\mu \sim 2$) singular isothermal sphere were found to satisfactorily reproduce the observed polarization pattern from the submm dust emission  at scales between $\sim300$ and 1500 au \citep{Goncalves08, Frau11}. However, at scales smaller than few hundreds au, the magnetic field departs from the idealized hourglass shape and becomes more perturbed, with hints of a toroidal field  \citep{Cox15, Liu16}.  In addition, this departure of the poloidal fields happens at the scales where IRAS 4A fragments in two subcores \citep[IRAS4A1, IRAS4A2 separated by $\sim 1\arcsec$ see][]{Santangelo15}. Although a complete kinematical study of IRAS4A is currently lacking to assess the presence of a disk at subarcsecond scales, we note that Ching et al. 2016 find a clear velocity gradient in the core at 1000 au scales, which is not detected at similar scales in B335, strengthening the scenario of a larger rotational torque in IRAS4A than in B335. 
Finally, in our first approach to model B335 we find that this source might have an initial higher mass-to-flux ratio than IRAS 4A, so the magnetic energy probably is overall weaker relative to the gravitational and rotational energies in IRAS4A than in B335.

In the L1157 star forming core, the magnetic field revealed from the polarized dust continuum at scales of $>500$ \citep{Stephens13} shows a very similar hourglass shape as the one observed in IRAS4A. However, contrary to IRAS4A, this source has no small-scale fragmentation, no large disk (upper limit set to a few tens of au) and no strong rotational motions are detected. 
This emphasizes the need of high angular resolution observations to robustly discriminate different flavors of magnetically-regulated collapse models, since an hourglass at large scales can arise from different  initial conditions, and outcomes of the protostellar accretion process. 
The B335 case could be more similar to the L1157 one if L1157 would be observed at similar physical scales (50 au), although our B335 ALMA map does not show a clear complete hourglass morphology at the envelope scale as the one observed in L1157. This is due to a combination of distance (L1157 is 2-3 times further than B335), and filtering of the extended emission from the lower column density outer envelope, in our ALMA observations.

L1527 in Taurus is another case of protostar seen almost perfectly edge-on, although it is probably more evolved than B335 \citep{Chen95,Ohashi96b}.
In L1527,  the 1.3 mm polarized continuum at scales of $500-1000$~au shows a magnetic field mostly aligned along the equatorial plane \citep{Hull14, SeguraCox15}, while at larger scales it is roughly aligned with cavity walls \citep{Davidson14}. This source has a larger disk radius, $\simeq$60 au, and rotation is detected \citep{Ohashi14}. Segura-Cox et al. (2015) use this example to suggest the topology of the B field lines could be a method to help identify candidate Class 0 disk sources with $R>10$ au because a transition from poloidal to radial field in the equatorial plane should be detected only in the sources where large rotational energy led to the formation of large disks: our observations of an equatorial component in B335, which harbors no disk at $R>10$ au, challenges this simple view, and we suggest both strong pinching and strong coiling of the field lines can  produce such a pattern.

We also stress that inclination effects (B335 is an isolated edge-on source) and a simple environment might (i) help to enhance the detection of the strongly pinched magnetic field in the equatorial plane, and (ii) identify whether the magnetic field is regulating the evolution of the collapsing matter.
For example, in more dynamic and/or clustered environments, polarization studies do not always reveal hourglass shapes, nor well-organized fields \citep{Zhang14}. Among the few other cases where the magnetic field has been resolved well enough, the Serpens protostar SMM1 observed with ALMA \citep{Hull17} shows a complex magnetic field morphology, although mostly aligned with the outflow's cavity. 


\section{Conclusions}

   \begin{enumerate}
      \item Our ALMA 233 GHz polarized dust emission map shows a well-organized pattern at scales 50--500 au, mostly distributed along two structures: the pronounced east-west outflow cavity walls where the field is aligned with the outflow direction and reminiscent of the large scale east-west native field, and the high-density equatorial plane where the field line segments appear to be mostly aligned with the equatorial plane. 
      \item Because observations suggest there is only room for very little rotation at $< 500$ au in B335, we suggest that the equatorial B-field component is created by strong pinching due to inward flow at these scales, consistent with the strong infall detected in the inner envelope of this protostar.
      \item A simple comparison of our magnetic field map to both numerical models of magnetized protostellar formation and the observed properties of B335 in the literature, suggest only a magnetically-regulated collapse scenario can explain the observed properties of B335 while reproducing the magnetic field topology observed with ALMA in this young Class 0 protostar.
   \end{enumerate}

\scriptsize{
\section*{Acknowledgements}
This paper makes use of the ALMA data ADS/JAO.ALMA$\#2013.1.01380.S$. ALMA is a partnership of ESO (representing its member states), NSF (USA) and NINS (Japan), together with NRC (Canada) and NSC and ASIAA (Taiwan) and KASI (Republic of Korea), in cooperation with the Republic of Chile. The Joint ALMA Observatory is operated by ESO, AUI/NRAO and NAOJ.
This project has received funding from the ERC under the European Union’s Horizon 2020 research and innovation programme (MagneticYSOS, grant agreement No 679937). 
J.M.G. acknowledges support from MICINN AYA2014-57369-C3-P (Spain).
S.P.L. acknowledges support from the Ministry of Science and Technology of Taiwan with grants MOST 105-2119-M-007-024 and MOST 106-2119-M-007-021-MY3.}



\bibliographystyle{mnras}
\input{ms.bbl} 

\bsp	
\label{lastpage}
\end{document}